\documentclass[eqsecnum,amsmath,preprintnumbers,superscriptaddress,nofootinbib,aps,11pt]
{revtex4} 
\usepackage{tikz}
\usetikzlibrary{calc,trees,positioning,arrows,chains,shapes.geometric,decorations.pathreplacing,decorations.pathmorphing,shapes,matrix,shapes.symbols}

\usepackage{bm}
\usepackage{MnSymbol}
\usepackage{amsmath,accents}
\usepackage{graphicx}
\usepackage{feynmp}
\usepackage{hyperref}
\usepackage{bbold}
\usepackage{eufrak}
\usepackage{upgreek,comment}
\usepackage{stmaryrd,scalerel}
\usepackage{mathtools}

\usepackage{enumitem}

\usepackage{color}
\usepackage{etex}
\usepackage{morefloats}

 \usepackage{longtable}

\makeatletter

\newcommand*{\overrightharpoonup}{\mathpalette{\overarrow@\rightharpoonupfill@}}
\newcommand*{\rightharpoonupfill@}{\arrowfill@\relbar\relbar\rightharpoonup}
\newcommand*{\overleftharpoonup}{\mathpalette{\overarrow@\leftharpoonupfill@}}
\newcommand*{\leftharpoonupfill@}{\arrowfill@\relbar\relbar\leftharpoonup}
\makeatother

\DeclareFontFamily{U}{matha}{\hyphenchar\font45}
\DeclareFontShape{U}{matha}{m}{n}{
      <5> <6> <7> <8> <9> <10> gen * matha
      <10.95> matha10 <12> <14.4> <17.28> <20.74> <24.88> matha12
      }{}
\DeclareSymbolFont{matha}{U}{matha}{m}{n}
\DeclareMathSymbol{\varrightharpoonup}{3}{matha}{"E1}

\def\bea{\arraycolsep .1em \begin{eqnarray}}
\def\eea{\end{eqnarray}}
\def\Tr{{\rm Tr}}

\def\!!!{\stackrel{!}{=}}
\def\???{\stackrel{?}{=}}

\def\a{\alpha}
\def\b{\beta}

\newcommand{\g}{ {\rm g} }

\def\eps{\epsilon}

\def\S{ \mathcal{S}}

\def\t{ \tau}

\def\n{ \Lambda }

\def\nn{ \nonumber \\}

\def\X{\hat{X}}

\def\x{{\hat{x}}}

\def\eq#1{(\ref{#1})}

\def\0#1#2{\frac{#1}{#2}}

\def\grgl{\:\hbox to -0.2pt{\lower2.5pt\hbox{$\sim$}\hss}{\raise3pt\hbox{$>$}}\:}
\def\klgl{\:\hbox to -0.2pt{\lower2.5pt\hbox{$\sim$}\hss}{\raise3pt\hbox{$<$}}\:}

\begin{document}
\title{Path integral measures and diffeomorphism invariance}

 \author{Alfio Bonanno}
 \address{\footnotesize{INAF, Osservatorio Astrofisico di Catania, Via S.Sofia 78, I-95123 Catania, Italy\\
 INFN, Sezione di Catania, Via S.Sofia 64, I-95123, Catania, Italy}}
\author{Kevin Falls}
\address{\footnotesize\mbox{Instituto de Física, Facultad de Ingeniería, Universidad de la República, J.H.y Reissig 565, 11300 Montevideo, Uruguay}}
	\author{Renata Ferrero}
	\address{\footnotesize\mbox{Institute for Quantum Gravity, FAU Erlangen – Nürnberg, Staudtstr. 7, 91058 Erlangen, Germany}}
	\vspace{1pt}
	\date{\today}

\date{\today}

\begin{abstract}
Much like the action, diffeomorphism invariance can be used to fix the form of the path integral measure in quantum gravity. Moreover, since there is a redundancy between what constitutes ``the action'' and what constitutes ``the measure'' one can always pick a minimal form of the latter. 
However, the authors of recent papers \cite{Branchina:2024lai,Branchina:2024xzh}  have advocated a form of the path integral measure for quantum gravity, proposed long ago by Fradkin and Vilkovisky, that is not invariant. This is easily seen since it depends explicitly on the $g^{00}$ component of the inverse metric without being contracted to form a scalar. An equally non-invariant measure was proposed in \cite{Donoghue:2020hoh}. 
As noted by their proponents, when these measures are used, certain divergences that typically appear are absent. However, the divergences that remain with the proposed measures are, unsurprisingly, neither diffeomorphism-invariant nor is the regulated effective action. We demonstrate this explicitly by computing the free scalar field contribution to the divergent part of the gravitational effective action using different measures and a proper-time cutoff. We support our findings with a thorough discussion of the path integral measure. In particular, we see how the contributions from the measure, obtained in a canonical setting, could be reinterpreted in a relational way compatible with diffeomorphism invariance.
\end{abstract}

\maketitle
\tableofcontents
\section{Introduction}

Reconciling quantum field theory (QFT) methods, which rely on renormalisation, with gravity has been a longstanding challenge. In recent years a search for a consistent theory has focused on the possibility that gravity is asymptotically safe \cite{Wetterich:1992yh,Morris:1993qb,Reuter:1996cp, Reuter:2019byg, Percacci:2017fkn,Reuter:2001ag,Bonanno:2020bil}. This approach is based on a path integral quantisation of general relativity combined with a Wilsonian approach to renormalisation.
Importantly, in contrast to other approaches, one does not write down a theory of quantum gravity, but instead searches the space of possible theories using the renormalisation group (RG) \cite{Manrique:2009tj,Manrique:2008zw,Falls:2017cze}.

An often overlooked ingredient of any path integral approach to quantum gravity is the path integral measure. In order to preserve BRST  (Becchi-Rouet-Stora-Tyutin) invariance at the quantum level the measure must be chosen to be invariant. Early attempts to define such a measure by Fradkin and Vilkovisky \cite{Fradkin:1973wke} include curious factors of the time-time component of the inverse metric $g^{00}$ arising after integrating to the momentum variables in the path integral. Explicitly there is a factor 
\begin{equation} \label{eq:factor}
\prod_x \left(\sqrt{g^{00}(x)}\right)^{N_{\rm dof}}\;,
\end{equation}
where $N_{\rm dof}$ is the number of propagating degrees of freedom  \cite{Unz:1985wq} (e.g. $N_{\rm dof}=2$ in pure gravity in four dimensions).
On the other hand a BRST invariant measure determined by Fujikawa \cite{Fujikawa:1983im,Fujikawa:2004cx} found no such factor $g^{00}(x)$. The situation was then partially clarified by Toms \cite{Toms:1986sh} who pointed out that compensating factors of $g^{00}$ must be included in the phase space integral in  order to get an invariant measure which agrees with Fujikawa's.
This makes sense since manifest invariance is lost when using phase space variables.
However, the situation is still to this day not uncontroversial  since Fradkin and Vilkovisky claim their measure is invariant \cite{Fradkin:1973wke} and offer a covariant version of it in \cite{Fradkin:1976xa}. In a Lorentzian setting the covariant version necessitates a ``physical time'' coordinate $\t(x)$. In particular, $\t(x)$ must depend on the physical degrees of freedom and should coincide with time in some coordinate systems. As we shall see, the covariant form is BRST invariant. On the other hand, it is not unique since there is no preferred physical time.

 It is important also to put formal statements into context of a specific regularisation which may itself be invariant or not. If dimensional regularisation is used the local factors of $g^{00}$ are typically disregarded, as taken as an isolated contribution to the effective action they give rise to a $\delta(0)$ singularity  which is set to zero in dimensional regularisation. On the other hand if one uses a covariant cutoff one has to take care to define the measure in a similarly covariant fashion in order to preserve BRST invariance.  Furthermore, even in dimensional regularisation one has to be aware of a possible  multiplicative anomaly which could lead  to different results depending on how the measure is combined at the level of determinants \cite{Ferrero:2023unz}.

Taking a fresh look at the situation in light of the asymptotic safety approach to quantum gravity, it is practical to view the measure and the action together as one structure. In this manner the search for a UV complete theory can concentrate on the action once an appropriately invariant measure is chosen. Indeed any modification to the measure which keeps it invariant can be always be rewritten as a modification of the action \cite{Manrique:2008zw,Manrique:2009tj}.\footnote{See \cite{Loll:1997iw,Dittrich:2008pw, Bahr:2011uj,Bahr:2009qc} for the discrete counterpart, where the discretization makes this construction even more fundamental.} Thus, from a practical point of view, it is more efficient to fix a minimal form of the invariant measure and consider general forms of the action. This minimal form is provided by Fujikawa's measure which prescribes the form of the measure for matter fields, gravity fluctuations and ghosts \cite{Fujikawa:1983im,Fujikawa:2004cx}.

The purpose of this paper is to address the controversy surrounding the form of the measure in view of some recent papers \cite{Branchina:2024xzh,Donoghue:2020hoh,Branchina:2024lai} where the effects of the measure have been discussed when cutoffs are used. For this purpose it is enough to concentrate on the measure for a free scalar field theory in curved spacetime as was the focus of \cite{Toms:1986sh}. Therefore, here we will have the same focus for most of the paper while summarising the situation for quantum gravity in the latter sections.

In section~\ref{sec:measure}, elaborating on  \cite{Toms:1986sh}, we give the minimal form of the  measure and demonstrate that it is invariant. In geometric terms the measure can be viewed as a volume element coming from an invariant DeWitt metric on the scalar fields configuration metric \cite{DeWitt:1957at,DeWitt:2003pm}. This formulation allows one to use different choice of variables provided the DeWitt metric is transformed. Using Fujikawa's ``preferred'' variables the volume element is a multiple of the identity. 
As we have mentioned it is important to prescribe how contributions determinants from the measure and Gaussian integrals are combined to avoid ambiguities.
This point is therefore briefly clarified.

One may want to derive the measure from the action or Hamiltonian \cite{Kuchar:1983rc}. This is possible if one has a measure on phase space as discussed in section~\ref{sec:PhaseSpace}. In ordinay quantum mechanics the measure is  (up to a  factor constant)  $\prod_t dq(t) dp(t)$. However, even in quantum mechanics, one can use a new time coordinate $x^0$ related to time by some invertible map $t = t(x_0)$. In this case using the variable $\phi(x^0) = q(t(x^0))$ one must transform the measure which entails a non-trivial factor if one uses $\phi(x^0)$ and its conjugates variable  $\pi(x^0)$ \cite{Toms:2004wp,Teitelboim:1981ua,Hartle:1986yu}. Similarly, in quantum field theory in flat spacetime one must use a non-trivial measure in phase space, if one uses curvilinear coordinates. For quantum field theory in curved spacetime and quantum gravity, things are less obvious since one does not have a preferred coordinate system \cite{Rovelli:2004tv,Rovelli:1990jm,Rovelli:1990pi,Rovelli:1995fv,Dittrich:2005kc,Goeller:2022rsx}.
Nonetheless using diffeomorphism-invariant field variables, which require a choice of some {\it physical} time parameter \cite{Thiemann:2007pyv,Thiemann:2004wk,Giesel:2007wn,Giesel:2012rb,Thiemann:2024vjx,Ferrero:2024rvi}, the trivial measure is by construction diffeomorphism invariant. In this manner the invariant version Fradkin and Vilkovisky's measure appears. However, once expressed in standard variables the analog of the $g^{00}$ factor is replaced by a ``Fradkin-Vilkovisky scalar'' depending on the physical time \cite{Fradkin:1976xa}. A minimal choice is to choose the physical time in order that this scalar is a constant. Then the path integral measure returns to the minimal form of Fujikawa.

In section~\ref{sec:Minimal} we evaluate the divergent part of the effective action for the minimal measure and recover the standard result. In particular finding that the divergent part is diffeomporphism invariant.   

In section~\ref{sec:General} we evaluate the divergent part of the effective action for a general local measure showing explicitly that if the measure is not invariant, the divergencies are not invariant either. It is explicitly shown that the local divergencies on a constant curvature spacetime are dependent on the coordinate system when the non-invariant  measure is used.

Section \ref{sec:brst} is dedicated to the revision of the BRST invariance of the measure in quantum gravity. In particular, we show how the residual freedom can be cast in the freedom to select a different action.
Importantly we clarify that, while in a gauge fixed setting it is not entirely wrong to have factors of $g^{00}$ in the measure, to preserve BRST invariance $g^{00}$ must be replaced by a the Fradkin-Vilkovisiky scalar $g^{\mu\nu} \partial_\mu \t \partial_\nu \t$ which coincides with $g^{00}$ when the gauge condition applies. Thus, once covariantised the contribution of the factor \eq{eq:factor} can be brought into the invariant action. Therefore, we once again recover the BRST minimal form of Fujikawa.

In section \ref{sec:flow} we  discuss briefly the implications of our findings for the RG flow in gravity.  
Without entering into a detailed calculation (since the effects are already visible for the case of a scalar) we summarise the main effects of choosing a measure and regularisation to the same effect as \cite{Branchina:2024lai,Branchina:2024xzh}. We explain in simple terms why this seems to disfavour a fixed point of the renormalisation group. However, the reality is that such choices break diffeomororphism invariance.  We note that an invariant measure and regularisation with the same consequences for the RG flow is possible. However, it is not unique and has a pathological flat spacetime limit.

Finally, in section \ref{sec:conclusion} we summarize and give an outlook. In particular, we conclude that, while there is much freedom to choose a measure, the conclusions of \cite{Branchina:2024lai,Branchina:2024xzh,Donoghue:2020hoh} should be revised.

In appendix \ref{app:details}
details can be found about the transformation of the heat kernel coefficients under a conformal transformation of the metric. This gives the steps we need to arrive at the key results of section~\ref{sec:General}.

\section{The invariant measure}
\label{sec:measure}
In this section, we introduce the key requirements for obtaining a diffeomorphism-invariant measure, highlighting the significance of the DeWitt covariant metric tensor and Fujikawa's variable. Additionally, we explore the relationship between the measure and the action, deriving a minimal prescription for the measure and discussing the impact of regularisation.

To support our argument, it is enough to consider the effective action we obtain by integrating out the degrees of freedom of a free minimally coupled scalar field in curved spacetime with a general ultra-local measure\footnote{In this paper we use the shorthand notation $\int_x = \int {\rm d}^dx$. Here and throughout the paper we use $g$ to denote the determinant of the metric tensor $g_{\mu\nu}$. However, note that in $
S[g]$, $\Omega_g$ etc. generally depend on the metric tensor $g_{\mu\nu}$ and its derivatives.} 
\begin{equation} \label{Gamma}
e^{-\Gamma[g]} = e^{-S[g]} \int D\left( g^{1/4} \Omega_g \phi\right) e^{-  \frac{1}{2} \int_x \sqrt{g} \left( g^{\mu\nu} \partial_\mu \phi \partial_\nu \phi + m^2 \phi^2 \right) }\;,
\end{equation}
where for any factor $M(x)$ we define
\begin{equation}
\int D\left( M \phi \right) \equiv \left(\prod_x \int d\phi(x) \frac{ M(x) }{\sqrt{2 \pi}} \right) \,.
\end{equation}
In  \eq{Gamma} $S[g]$ is a purely gravitational action and we refer to $\Omega_g(x)$ as the {\it measure factor}.
The factor $g^{1/4}$, which we pull out from the definition of the measure factor, is conventional. We assume that $\Omega_g(x)$ can depend on the metric and its derivatives evaluated at $x$.

Our aim is to find {\it an} expression for $\Omega_g(x)$ under the requirement that $\Gamma[g]$ is invariant under diffeomorphisms of the metric. 
Specifically, we demand that  
\begin{equation}
\Gamma[g'] = \Gamma[g] 
\end{equation}
with 
\begin{equation}
g_{\mu\nu}'(x) = g_{\rho\lambda}(\hat{X}(x)) \partial_\mu \hat{X}^\rho(x) \partial_\nu \hat{X}^\lambda(x)\,,
\end{equation}
where $\hat{X}(x)$ is an invertible map, with the inverse $X(\hat{x})$, such that $\hat{X}(X(\hat{x})) = \hat{x}$ and $X(\hat{X}(x)) = x$.  

Other than invariance, in this section, we will opt for simplicity to determine a ``minimal'' form for $\Omega_g$. We will then demonstrate that a different action $S[g]$ can equivalently incorporate more involved, ``non-minimal'' choices for $\Omega_g$.

 The action for the scalar field is diffeomorphism invariant under transformations which act both on the metric and on the scalar field, with
\begin{equation}
     \phi'(x) = \phi(\hat{X}(x))\;.
\end{equation}
Therefore $\Gamma[g]$ will be invariant if  
\begin{equation}
 D(g'^{1/4} \Omega_{g'} \phi') =  D(g^{1/4} \Omega_g \phi)\,.
\end{equation}
An elegant geometric manner to define such a measure is to write a diffeomorphism invariant DeWitt metric $\delta \ell^2 $ on the configuration space of the scalar field and then identify $ D(g^{1/4} \Omega_g \phi)$ with the corresponding volume element. 
It is easy to find such a metric. Indeed a simple invariant metric is
\begin{equation} 
\delta \ell^2 =  \int_x \mu^2  \sqrt{g}(x)\delta \phi(x) \delta \phi(x) \,.
\end{equation}
The corresponding  covariant DeWitt metric tensor is given by\footnote{Here and throughout $\delta(x-y)$ is the standard delta function $\int_y \delta(x-y) f(y) = f(x)$.  }
\begin{equation} \label{Scalar_DeWitt_Metric}
\gamma_\phi(x,y) =  \mu^2 \sqrt{g}(x) \delta(x-y)\,.
\end{equation}
Here $\mu$ is a constant that should have mass dimension one such that the measure is dimensionless. The metric \eq{Scalar_DeWitt_Metric} is ultra-local meaning that it is proportional to a delta-function.  
Following our prescription an invariant measure is given by 
\begin{equation} \label{Simple_Measure}
D(g^{1/4} \Omega_g \phi)= D\phi \sqrt{\det \gamma_\phi[g]} =  D(g^{1/4} \mu \phi)\,,\,\,\,\,\,\,\,\,\,\, D\phi \equiv \left(\prod_x   \frac{d\phi(x)}{\sqrt{2 \pi}} \right)  \,,
\end{equation}
thus evaluating the determinant yields
 \begin{equation} \label{Omega_mu}
\Omega_g = \mu\;.
 \end{equation}
Hence, we learn that writing down an invariant measure is no more difficult than writing down an invariant action.

 \subsection{Diffeomorphism invariance using Fujikawa's variable}
In this subsection, we want to  show that the measure \eq{Simple_Measure} is diffeomorphism invariant.
To this end, let us note that we can also write
\begin{equation}
  D(\phi) \sqrt{\det \gamma_\phi[g]} = D( \mu  \phi_F) \;,
\end{equation}
where we define
\begin{equation}
\phi_F \equiv g^{1/4}\phi
\end{equation}
 which is  Fujikawa's variable \cite{Fujikawa:1983im,Fujikawa:2004cx}. This represents  a density of weight one half. In terms of $\phi_F$ the effective action is
\begin{equation} \label{GammaF}
e^{-\Gamma[g]} = e^{-S[g]} \int D(\mu \phi_F) e^{-  \frac{1}{2} \int_x \left( g^{\mu\nu} \nabla_\mu \phi_F \nabla_\nu \phi_F + m^2 \phi^2_F \right) }\;,
\end{equation}
where we note that the covariant derivative of $\phi_F$ is given by $\nabla_\mu \phi_F = \partial_\mu \phi_F- \frac{1}{4} g^{-1}\partial_\mu g \phi_F$. 
 Fujikawa's variable  transforms as 
\begin{equation}
\phi_F(x) \to  \int_{\hat{x}} P(x,\hat x)   \phi_F(\hat{x})
\end{equation}
with $P$ given by 
\begin{equation}
P(x,\hat{x} ) =  \sqrt{\det \frac{\partial \hat X }{\partial x } } \delta(\hat{X}(x)-\hat{x} ) \,.
\end{equation}
We remind the reader that  $\hat{X}^\mu(x)$ is a diffeomorphism, i.e., an invertible map from the manifold to itself. The inverse of  $\X^{\mu}(x)$ is then simply $X^{\mu}(\hat{x})$.
It follows that 
\begin{equation}
    \delta(\hat{X}(x),\hat{x} ) =  \det \frac{\partial  X }{\partial \hat{x} } \delta(x-X(\hat{x})) 
\end{equation}
and 
\begin{equation}
  \det \frac{\partial \hat X }{\partial x } = \frac{1}{\det \frac{\partial  X }{\partial \hat{x} }}  \;.
\end{equation}
Then, alternatively, using the last two identities, $P$ can also be written as 
\begin{equation}
P(x,\hat{x} )  =    \sqrt{\det \frac{\partial  X }{\partial \hat{x} } } \delta(x - X(\hat{x})) \,.
\end{equation}

From these expressions, we see that $P(x,\hat{x} )$ is orthogonal (its transpose is its inverse) and hence $\det P =1$.
Therefore the measure is invariant
\begin{equation}
  D( \mu  \phi_F)  \to  D( \mu  \phi_F)  \det P = D( \mu  \phi_F) \;,
\end{equation}
as we required.

\subsection{Diffeomorphism invariance with the DeWitt metric}
We can also show the invariance of the action using the transformation of the DeWitt metric \cite{DeWitt:1957at,DeWitt:2003pm}.
It is not hard to show that the DeWitt metric transforms as 
\begin{equation}
\gamma'_\phi(x,y) =  \int_{\hat{x},\hat{y}}    \gamma_\phi(\hat{x},\hat{y})    \delta(X(\hat{x})-x) \delta(X(\hat{y})-y) \;,
\end{equation}
while $\phi(x)$ transforms like
\begin{equation}
\phi(x) \to \int_{\hat{x}}  \delta(\hat{X}(x)-\hat{x}) \phi(\hat{x})\;.
\end{equation}
Defining
\begin{equation}
T(x,\hat{x}) =  \delta(\hat{X}(x)-\hat{x}) 
\end{equation}
and
\begin{equation}\label{eq:hatT}
\hat{T}(\hat{x},x)  = \delta(X(\hat{x})-x) 
\end{equation}
one easily sees that these are inverses, namely that their integrated product gives a delta function
\begin{equation}
\int_x \hat{T}(\hat{x},x) T(x,\hat{y}) = \delta(\hat{x}-\hat{y})\;.
\end{equation}
Thus, we see again that the measure is invariant under a diffeomorphism transformation:
\begin{equation}
 D\phi \sqrt{\det \gamma_\phi[g]} \to  \det  D\phi \det T  \sqrt{\det \gamma_\phi[g]} \det \hat{T} = D\phi  \sqrt{\det \gamma_\phi[g]}\;.
\end{equation}

\subsection{Generality}

One may complain that $\sqrt{\det \gamma_\phi[g]}$ is not unique and ultimately the correct measure can be found only by comparing with experiment. 
However, if we allow for the freedom to write down {\it any} action then any modification to the measure can be rewritten as 
\begin{equation}
\sqrt{\det G_\phi} e^{-S} = \sqrt{\det \gamma_\phi} e^{-S'}\;
\end{equation}
for a new action $S'= S - \Tr \log G_\phi \gamma^{-1}_\phi$. Furthermore, $S'$ will remain invariant provided $G_\phi$ transforms as $\gamma_\phi$ does. Thus, if we suppose the actions $\Gamma[g]$ and $S[g]$ are diffeomorphism invariant, but otherwise general, without losing generality we can fix the measure to the simple form \eq{Simple_Measure} and limit our freedom to construct different theories to the action.
Indeed, the most general theory can be written down such that it depends on a single functional $\rho[\phi,g]$ such that
\begin{equation}
{\rm e}^{-\Gamma[g]} = \int D\phi \, \rho[\phi,g]\;,
\end{equation}
where we require that
\begin{equation} \label{rho_transform}
\rho[\phi',g'] =  \rho[\phi,g]  \det \hat{T}\;
\end{equation}
with $\hat T$ defined in \eqref{eq:hatT}.
Then, we can define by a convention that the action $S[\phi,g]$ is
\begin{equation}
e^{-S[\phi,g]} \equiv  \frac{1}{\sqrt{\det \gamma_\phi[g]}}  \rho[\phi,g]\;,
\end{equation}
where we allow now for the most general action $S[\phi,g]$.
Other conventions can be more or less useful depending on the form of $\rho[\phi,g]$  but ultimately there is no ``correct'' way to distinguish the measure from the action \cite{Manrique:2008zw,Manrique:2009tj}.
What does matter is the transformation law \eq{rho_transform}. 

\subsection{Multiplicative Ambiguities and Anomalies}
One subtlety that can arise is when regulated determinants are such that
\begin{equation}
 \sqrt{\det (A B)} = \sqrt{\det A} \sqrt{\det B} +\mathcal{A}[A,B]
\end{equation}
where $A$ and $B$ are two operators and $\mathcal{A}[A,B]$ is a multiplicative ambiguity which could in the worse case break diffeomorphism invariance.
In this case, one should be careful to combine determinants such that the regularised determinants contributing to the effective action are finite and diffeomorphism invariant.  To avoid any ambiguity  we can introduce an inner product  
\begin{equation}
(\psi , \phi) =  \int_x \sqrt{g} \psi \phi 
\end{equation}
and then {\it define} the measure by 
\begin{equation} \label{MeasureDef}
\int d(g^{1/4} \Omega_g \phi) e^{-\frac{1}{2}(\phi ,\Delta \phi) } := \sqrt{\det \Delta/\Omega^2_g} 
\end{equation}
for any differential operator $\Delta$. 
This removes any ambiguity but must be revisited when the determinant is regularised to again give an unambiguous meaning the regularised integral.

\section{Phase space}
\label{sec:PhaseSpace}
We now focus on deriving the measure from phase space integration. To do so we generalise the procedure of quantisation in quantum mechanics to quantum field theory in curved spacetime. This can be achieved in three steps. First, we consider the path integral in standard quantum mechanics and rewrite it in a time reparameterisation invariant way, i.e. so that it is invariant under one dimensional diffeomorphisms \cite{Teitelboim:1981ua,Hartle:1986yu,Bahr:2011uj}. This already reveals that one has to be careful to define the phase space measure of the path integral. In the second step we generalise the first step by treating QFT in flat spacetime in an arbitrary curvi-linear coordinate system. To reach the measure in curved spacetime, in a third step, we then simply use Einstein's minimal coupling prescription applied to the measure. 

These considerations fix the measure up to non-minimal terms, analogous to a curvature dependent term in the action of the scalar $ \frac{\xi}{2} \int_x\sqrt{-g} R(\phi^2)$.

\subsection{Time reparameterisation invariant quantum mechanics}
In this section we consider a simple harmonic oscillator in quantum mechanics. The action is given by
\begin{equation}\label{eq:3.1}
S[q]= \int dt L =\int dt\left( \frac{1}{2} \dot{q}(t)^2 - \frac{1}{2} m^2 {q}(t)^2 \right)\;.
\end{equation}
From the Lagrangian $L$ we obtain the canonical momentum in the standard way, finding  
\begin{equation}
   p(t) =  \dot{q}(t)\;.
\end{equation}
This then leads us to the Hamiltonian 
\begin{equation} \label{HQM}
{\rm H} =  \frac{1}{2} p(t)^2 + \frac{1}{2} m^2 q(t)^2\,. 
\end{equation}
The path integral written as an integral over phase space is given by the integration of configuration space and the conjugate momenta. The generating functional is then
\begin{equation} \label{QM_pathintegral}
Z= \int D(\mu q) \int Dp   e^{{\rm i} \int  dt (\dot{p} q -{\rm H})}\,.
\end{equation}
To make the connection with the QFT path integral, one can integrate out the momenta $p$. This  amounts to recover the standard result
\begin{equation} \label{ZAction}
Z =  \int D(\mu q)   e^{{\rm i}S[q] }\,.
\end{equation}

Now we want to work in a relativistic setting. In full generality, we need to introduce a general time coordinate $x^0$ in order to express the proper time interval  as 
\begin{equation}
dt^2 =g_{00}(x) dx^0dx^0   \;, 
\end{equation}
where $g_{00}(x)$ is the one-dimensional metric. From this it follows that
\begin{equation}
\frac{dt}{dx^0} = \sqrt{g_{00}} \implies  \frac{dx^0}{dt} = \frac{1}{\sqrt{g_{00}}}\;.
\end{equation}
Furthermore, to establish the link with field theory more manifestly, we introduce a variable\footnote{In this subsection $\phi(x)$ is shorthand for $\phi(x^0)$, since we are working with a  ``$1+0$ dimensional field theory''.} $\phi(x)$, as an  alternative variable to $q(t)$, by defining
\begin{equation}
    \phi(x) = q(t(x)) \implies q(t) = q(x(t))\;.
\end{equation}
Thus, $t(x)$ is a one dimensional diffeomorphism or time reparameterisation.
In one dimension we note that the determinant is $g = g_{00}$ and that the inverse metric is $g^{00} = 1/g_{00}$.

In view of that, in a relativistic setting the action \eq{eq:3.1} can be rewritten as 
\begin{equation}
 S[\phi,g] = \int dx \sqrt{g} \left(\frac{1}{2} g^{00} \partial_0 \phi(x) \partial_0 \phi(x)  - \frac{1}{2} m^2 \phi(x)^2\right)\;.
\end{equation}

We should now be able to  define the generating functional $Z$ in a covariant manner such that it is independent of the coordinate system and hence $g_{00}$. This must then involve a quantisation procedure that preserves time reparameterisation invariance \cite{Bahr:2011uj}. Let us note that, the measure \eq{Simple_Measure} with $\Omega_g = \mu$ is invariant under diffeomorphisms, $q$ and $\phi$ are related by a diffeomorphism and that in the coordinate system where $x(t)=t$ we have $g \!!! 1$. 
It follows  that
\begin{equation}
  D(\mu g^{1/4} \phi) = D(\mu q)  \,.
\end{equation}
 Thus, directly from the standard path integral \eqref{ZAction} we obtain 
\begin{equation} \label{Z_final}
    Z = \int  D(\mu g^{1/4} \phi) {\rm e}^{{\rm i} S[\phi,g]}\,.
\end{equation}
Alternatively, one could start from the phase space formulation, defining as usual 
\begin{equation}
 \pi = \frac{\partial L}{\partial (\partial_0 \phi)} = \sqrt{g} g^{00} \partial_0 \phi \,.
 \end{equation}
At this stage, it is important to emphasize that there is a difference between the integration differentials coming from the quantum mechanical variables, namely $D \phi D \pi \neq Dq D p$. On the other hand one has that
\begin{equation}
 D \phi  D \left( \left(g^{00}\right)^{-1/2} \pi \right)   =  Dq D p\;,
\end{equation}
which follows from the transformation of $\left(g^{00}\right)^{-1/2} \pi$  as a density of weight one. We can write then
\begin{equation}
 D \phi  D \left( \left(g^{00}\right)^{-1/2} \pi \right) =D( g^{1/4} \phi)  D \left( g^{-1/4} \left(g^{00}\right)^{-1/2} \pi \right)  \;,
\end{equation}
so that both $D( g^{1/4} \phi)$ and $D\left( g^{-1/4} \left(g^{00}\right)^{-1/2} \pi \right)$   are invariant under time reparameterisations. Thus, separately the identifications would read
\begin{equation}
D( g^{1/4} \phi)  = Dq
\end{equation}
and
\begin{equation}
    D\left( g^{-1/4} \left(g^{00}\right)^{-1/2} \pi \right) =  Dp\;.
\end{equation}
Note that $\phi(x)$ and $\pi(x)$ are not the canonical coordinates on phase space since these are $q(t)$ $p(t)$. Instead, we should view $\phi(x)$ and $\pi(x)$ as alternative variables on phase space which use a different notation of time namely the coordinate time $x^0$ instead of the proper time $t$.
 The Hamiltonian in these alternative variables is
\begin{equation} \label{HQMphipi}
    H =\frac{1}{2} \frac{1}{ g^{00} \sqrt{g}} \pi(x)^2 +  m^2 \sqrt{g} \phi(x)^2 \,.
\end{equation}
Then we can rewrite the phase space generating functional \eq{QM_pathintegral} as
\begin{equation}
    Z = \int D(\mu \phi) D \left( \left(g^{00}\right)^{-1/2} \pi \right) e^{i\int d x ( \pi(x) \partial_0 \phi(x)- H(x)) }\;.
\end{equation}
Crucially, integrating out the momenta we obtain again \eq{Z_final}. This teaches us an important lesson \cite{Toms:1986sh}: in order to get the correct measure in standard quantum mechanics we must use the right phase space measure which is determined by using the canonical variables $q$ and $p$ first with the standard measure on phase space. Then we can re-express the path integral in terms of other variables taking care to transform the measure. If we instead use the phase space measure  $\int D\phi \int D\pi$ we would naively get
\begin{equation}
    Z = \int  D(g^{00} g^{1/4} \phi) e^{{\rm i} S[\phi]}\;,
\end{equation}
which is incorrect.

\subsection{QFT in flat spacetime}
As a next step, we will generalise the procedure applied in quantum mechanics to QFT in flat spacetime in an arbitrary coordinate system. Let us use $\psi(\hat{x})$ to denote the scalar field in the usual Minkowski coordinates where $g_{\mu\nu}(\hat{x}) = \eta_{\mu\nu}$. Then the action reads
\begin{equation}
S_{\eta}[\psi] =-\frac{1}{2} \int_{\hat{x}} ( \eta^{\mu\nu} \partial_\mu \psi \partial_\nu \psi + m^2 \psi^2)\;.
\end{equation}
We define the canonical momentum as 
\begin{equation}
\varpi =   \frac{\partial \mathcal{L} }{ \partial \partial_0\psi} = \partial_0 \psi\;.
\end{equation}
Then the canonical form of the Hamiltonian density is then
\begin{equation}
{\rm H}(\psi, \varpi) = \frac{1}{2} \varpi^2  +  \frac{1}{2} \delta^{ij} \partial_i \psi \partial_j \psi + \frac{1}{2} m^2 \psi^2\;.
\end{equation}
This represents the analog of the Hamiltonian \eq{HQM} in quantum mechanics. 
We can then go from the Hamiltonian form of the generating functional
\begin{equation} \label{Z_QFT}
    Z = \int  D(\mu \psi) D(\varpi)  e^{{\rm i}\int_x ( \partial_0\varphi \varpi - {\rm H}(\varphi,\varpi, \g)) }
\end{equation}
to the action form as in quantum mechanics obtaining 
\begin{equation} \label{Z_QFT}
    Z = \int  D(\mu \psi) e^{{\rm i} S_\eta[\psi]}\;.
\end{equation}
At this stage, we promote the field $\phi(x)$ to be the field in an arbitrary coordinate system such that it transforms
\begin{equation}
    \phi(x) = \psi(\hat{X}(x)), \,\,\,\,   \psi(\x) = \psi(X(\x))
\end{equation}
under some diffeomorphism $\hat{X}(x)$.
An arbitrary flat metric is then given by
\begin{equation}
g_{\mu\nu}(x) = \eta_{\rho\lambda} \partial_\mu \hat{X}^\rho(x) \partial_\nu \hat{X}^\lambda(x)\,.
\end{equation}
The standard measure can be rewritten as
\begin{equation}
D(\mu \psi) = D(\mu (-\eta)^{1/4} \psi)=  D(\mu (-g)^{1/4} \phi)
\end{equation}
where we used  that the determinant is $-\eta =1$  and  the diffeomorphism invariance of the measure. Furthermore, the action  written in terms of $\phi$ and a general coordinate system $g_{\mu\nu}$ is then 
\begin{equation}
S[\phi,g] = - \frac{1}{2} \int_x \sqrt{-g} ( g^{\mu\nu} \partial_\mu \phi \partial_\nu \phi + m^2 \phi^2)\;.
\end{equation}
Again, let us emphasize that this is equal to $S_\eta[\psi]$ by diffeomorphism invariance. So going from  the path integral in flat spacetime \eq{Z_QFT} to the one  in an arbitrary coordinate system amounts to
\begin{equation} \label{Z_final_QFT}
    Z = \int  D\left(\mu  (-g)^{1/4} \phi\right) {\rm e}^{{\rm i} S[\phi,g]}\,.
\end{equation}
Let us now analyze what is happening at the level of the phase space. 
 The alternative phase space momentum coordinate is
\begin{equation}
\pi =  \frac{\partial \mathcal{L} }{ \partial( \partial_0 \phi)} =- \sqrt{- g} g^{0\nu}\partial_\nu \phi\;.
\end{equation}
In terms of these alternative variables the Hamiltonian is then given by 
\begin{equation}
H(\phi,\pi, g) =  \frac{1}{2}\frac{1}{\sqrt{g} (-g^{00})} \pi^2 + \frac{1}{\sqrt{- g^{00} }}\pi g^{0i} \partial_i \phi + \frac{1}{2} \sqrt{g} g^{0i} \partial_i \phi g^{0j} \partial_j \phi + \frac{1}{2} \sqrt{g} g^{ij} \partial_i \phi \partial_j \phi\,.
\end{equation}
Considering then the phase space integral. This is given by
\begin{equation}
Z= \int D(\mu \phi) D\left( \Xi  \pi \right)   e^{{\rm i}\int_x (\partial_0\phi \pi -  H(\phi,\pi, g)) } \;,
\end{equation}
where we will determine $\Xi$ by reproducing \eq{Z_final_QFT}.
Finally, integrating over $\pi$ we obtain
\begin{equation} \label{Z_Xi}
    Z = \int  D\left(\mu \sqrt{-g^{00}} \Xi (-g)^{1/4} \phi\right)  {\rm e}^{{\rm i} S[\phi,g]}\;.
    \end{equation}
Hence we can determine that $\Xi = 1/\sqrt{- g^{00}}$. 
Therefore, we find that the correct phase space integral in terms of the alternative variables in flat spacetime is
\begin{equation} \label{ZQFT_H}
Z= \int D(\mu \phi) D\left( \left( g^{00}\right)^{-1/2}   \pi \right)   e^{{\rm i}\int_x (\partial_0\phi \pi - {\rm H}(\phi,\pi, g)) }\,.
\end{equation}

\subsection{QFT in curved spacetime}
 By the minimal coupling principle we can take the measure of flat spacetime written in terms of a general $g_{\mu\nu}$ and simply generalise it to curved spacetime \cite{Parker:1979mf,Parker:2009uva}. Therefore we now take $g_{\mu\nu}$ to be an arbitrary curved spacetime metric and we allow for a pure gravity action $S[g]$. Then \eq{ZQFT_H} generalises to  
 \begin{equation}
e^{{\rm{i}}\Gamma[g]} = \int D(\mu \phi) D\left( \left( g^{00}\right)^{-1/2}  \pi \right)   e^{{\rm i}\int_x (\partial_0\phi \pi -  H[\phi,\pi, g]) } e^{{\rm{i}} S[g]}
\end{equation}
Integrating out the momenta, $\pi$ we obtain 
\begin{equation} \label{Gamma_final_QFT}
    e^{{\rm{i}}\Gamma[g]}  = \int  D\left(\mu  (-g)^{1/4} \phi\right) {\rm e}^{{\rm i} S[\phi,g]}\,,
\end{equation}
where
\begin{equation}
     S[\phi,g] = - \frac{1}{2} \int_x \sqrt{-g} ( g^{\mu\nu} \partial_\mu \phi \partial_\nu \phi + m^2 \phi^2) + S[g]\;.
\end{equation}
We note that if we rotate to Euclidean signature we get our previous result.\footnote{The Wick rotation procedure could represent a delicate step, for further discussions we refer the reader to \cite{Baldazzi:2018mtl}.} In this way we obtain the minimal measure from the flat spacetime expression exactly in the same manner as it is done for the action.
One could add non-minimal terms, which, however,  can be absorbed into the action. 

As an alternative to the minimal coupling principle, we can also require that for some choices of variables $\varphi$ and $\varpi$ the quantisation rule is canonical, meaning that the measure is simply 
$ \int D(\mu \varphi) D \varpi$. This can be achieved by   relating $\varphi$ to $\phi$ by a diffeomorphism $\phi(x) = \varphi(\X(x))$ and then requiring that the corresponding metric $\g_{\mu\nu}$ satisfies $\g^{00} = -1$. Then the corresponding Hamiltonian is  
\begin{equation}
{\rm H}[\varphi,\varpi, \g] =  \frac{1}{2}\frac{1}{\sqrt{\g}} \varpi^2 + \varpi \g^{0i} \partial_i \varphi + \frac{1}{2} \sqrt{\g} \g^{0i} \partial_i \phi \g^{0j} \partial_j \varphi + \frac{1}{2} \sqrt{\g} \g^{ij} \partial_i \varphi \partial_j \varphi\,.
\end{equation}
Here $\varpi$ is the momentum conjugate to $\partial_0 \varphi$. Then we obtain by integrating out $\varpi$
\begin{equation} \label{Gamma_final_varphi}
    {\rm e}^{ {\rm i}   \Gamma[\g]}  = \int  D\left(\mu  (-\g)^{1/4} \varphi\right) {\rm e}^{{\rm i} S[\varphi,\g]}\,,
\end{equation}
where, by the diffeomorphism invariance of the measure, we have 
\begin{equation}
    \int  D\left(\mu  (-\g)^{1/4} \varphi\right) {\rm e}^{{\rm i} S[\varphi,\g]} = \int  D\left(\mu  (-g)^{1/4} \phi\right) {\rm e}^{{\rm i} S[\phi,g]} = e^{{\rm i}\Gamma[g]}\;.
\end{equation}
Hence,  we recover \eq{Gamma_final_QFT}.

We conclude that there are not a unique set of variables to use on phase space and thus, even for quantum mechanics, one must use the canonical variables to be sure that the phase space measure is the standard one (i.e., with $Dp Dq$ up to a constant factor). In curved spacetime one can impose that the ``canonical variables'' are those such that the corresponding metric satisfies $\g^{00} = -1$ which generalises the quantum mechanical case and agrees with the minimal coupling principle. 

\subsection{Relational quantisation}
One might not be satisfied with the minimal coupling procedure or the choice to fix the canonical variables to those associated to the metric with $\g^{00} = -1$. Another option to fix the measure is to choose to use a relational field $\Phi(\sigma)$, which is diffeomorphism invariant, and quantise using this variable with a standard measure on phase space $D(\mu \Phi) D(\Pi)$. This step is at the core of the reduced phase space quantisation program \cite{Thiemann:2004wk, Giesel:2007wn,Giesel:2012rb,Ferrero:2024rvi}.

To achieve this, we have to construct $d$ scalars $\Sigma^{I}(x)$ (with $I = 0,1,2, ..,d$) from the metric (and/or other fields) which are diffeomorphisms  with inverse $X^\mu(\sigma)$ \cite{Brunetti:2016hgw,Frob:2017gyj,Frob:2017lnt,Baldazzi:2021fye}.
The fields $\Sigma^{I}(x)$ must be composite fields that transform as scalars and constitute physical coordinates. In particular
\begin{equation}
\t (x) \equiv  \Sigma^0(x) 
\end{equation}
is understood as a physical time parameter.
This being the case, one has that
\begin{equation}
\Phi(\sigma) = \phi(X(\sigma))
\end{equation}
is diffeomorphism invariant and we similarly get a diffeomorphism invariant metric $G_{IJ}(\sigma)$ and a momentum variable $\Pi(\sigma)$.
Here the coordinates $\sigma^I$ represent the values of the scalar field $\Sigma^I$ so $X^\mu(\sigma)$ are the coordinates $x^\mu = X^\mu(\sigma)$ 
of the points where $\Sigma^I = \sigma^I$. In this way the coordinates $\sigma^I$ have a physical meaning (for some examples see \cite{Brown:1994py,Giesel:2012rb} for  a realisation with  dust physical coordinates).

There is a huge freedom in the choice of such scalars \cite{Gambini:2000ht}. However, let us require i) that the measure has the correct flat spacetime limit ii) that the measure factor $\Omega_g$ can be expanded in derivatives of the metric. That is we can write
\begin{equation}\label{eq:3.42}
\Omega_g  = \mu\left(a_0 + a_1 R/\mu^2 + a_{2a} R^2/\mu^4 + a_{2b}  R_{\mu\nu}R^{\mu\nu}/\mu^4 + \nabla^2R a_{2c} + a_{2d} R_{\mu\nu\sigma \lambda} R^{\mu\nu\sigma \lambda}/\mu^4 + \dots \right)\;,
\end{equation}
where we should put $a_0=1$ to get the flat spacetime limit.
Computing the path integral for general $\Sigma^I$ we find 
\begin{equation} \label{Z_rel}
    Z = \int  D\left(\mu \sqrt{-G^{00}}  (-G)^{1/4} \Phi\right)  {\rm e}^{{\rm i} S[\Phi,G]}\;.
    \end{equation}
The factor $\mu \sqrt{-G^{00}}  (-G)^{1/4} $ can be written as
\begin{eqnarray}
    \sqrt{-G^{00}}  (-G)^{1/4}(\sigma)  &=& \sqrt{-g^{\mu\nu}  \partial_\mu \t \partial_\nu \t} (-\det e_I^\mu e_J^\nu g_{\mu\nu})^{1/4}|_{x = X(\sigma)}\nn
    &=& \left. \sqrt{-g^{\mu\nu}  \partial_\mu \t \partial_\nu \t} (- g)^{1/4} e^{1/2}\right|_{x= X(\sigma)}\;,
\end{eqnarray}
where $e^\mu_I(x)$ is the inverse of matrix of partial derivatives of the fields $\partial_\mu \Sigma^I(x)$ and $e = \det e^\mu_I$. 

We now want to transform the measure to a form in terms of the standard diffeomorphism variant fields $\phi$ and $g_{\mu\nu}$. To this end, let us write the DeWitt metric tensor in the  physical coordinates
\begin{equation}
    \upgamma_\Phi(\sigma, \tau) = \left. -\mu^2 e \sqrt{-g}   g^{\mu\nu}  \partial_\mu \t \partial_\nu \t \right|_{x= X(\sigma)} \delta(\sigma - \tau)\;.
\end{equation}
  Then we can write the measure in \eqref{Z_rel} in the geometric form 
  \begin{equation}
 D\left(\mu \sqrt{-G^{00}}  (-G)^{1/4} \Phi\right) =  D\Phi \sqrt{\det  \upgamma_\Phi } \;.
  \end{equation}
  The corresponding metric line element is given by
\begin{eqnarray}
\delta \ell^2 &=&   \left.     -\mu^2\int {\rm d}^d \sigma  \, e \sqrt{-g}   g^{\mu\nu}  \partial_\mu \t \partial_\nu \t  \right|_{x= X(\sigma)} \, \delta \Phi(\sigma)^2 \nn
&&= \left. -\mu^2\int {\rm d}^d \sigma  \, e \sqrt{-g}   g^{\mu\nu}  \partial_\mu \t \partial_\nu \t  \right|_{x= X(\sigma)} \, \delta \phi(X(\sigma))^2\nn
&&=  -\mu^2\int_x \sqrt{-g}   g^{\mu\nu}  \partial_\mu \t \partial_\nu \t \,  \delta \phi(x)^2  \;.
\end{eqnarray}
From this, we can read off the DeWitt metric tensor in the $\phi$ variables:
\begin{equation}
    \upgamma_\phi(x, y) = -\mu^2 \sqrt{-g}   g^{\mu\nu}  \partial_\mu \t \partial_\nu \t  \delta(x -y)
\end{equation}
and finally we can write the measure
\begin{equation} \label{InvariantFV}
D\Phi \sqrt{\det  \upgamma_\Phi } = D\phi \sqrt{\det  \upgamma_\phi } = D\left(\mu \sqrt{- g^{\mu\nu}  \partial_\mu \t \partial_\nu \t}    (-g)^{1/4}\phi \right)\;.
\end{equation}  
Importantly, this corresponds to a measure factor
\begin{equation}
\Omega_g = \mu \sqrt{- g^{\mu\nu}  \partial_\mu \t \partial_\nu \t}\;.
\end{equation}
We emphasize that this is diffeomorphism invariant since $\sqrt{-g^{\mu\nu}  \partial_\mu \t \partial_\nu \t}$ 
is a scalar. The measure \eq{InvariantFV} is of the same form as the measure used in \cite{Fradkin:1976xa} and is the proper covariant version of the measure derived in \cite{Fradkin:1973wke}. However, it is important to stress that here $\t$ is a scalar playing the role of a physical time coordinate in a Lorentzian framework. There are many such physical times on a generic spacetime.

On the other hand, our requirements for \eqref{eq:3.42} lead to
\begin{equation} \label{Curvauture_expansion}
- g^{\mu\nu}  \partial_\mu \t \partial_\nu \t = 1 + a_1 R/\mu^2 + a_{2a} R^2/\mu^4 + a_{2b}  R_{\mu\nu}R^{\mu\nu}/\mu^4 + \nabla^2R a_{2c} + a_{2d} R_{\mu\nu\sigma \lambda} R^{\mu\nu\sigma \lambda}/\mu^4 + \dots \,.
\end{equation}
This is not possible for any local scalar field $\t$ constructed from the metric (or any other field). Now, although there are many choices for $\t$, they must lead to \eq{Curvauture_expansion} for some fixed values of the coefficients. Thus, whatever such choice of $\t$ is made, it will be equivalent to demanding \eqref{Curvauture_expansion} to hold for some values of the coefficients appearing on the RHS. This amounts then to solve a differential equation for $\t$. With appropriate covariant boundary conditions, we can then determine $\t$ depending on the coefficients.

For instance, writing $g_{\mu\nu} =\eta_{\mu\nu} + h_{\mu\nu}$
we can determine $\t$ as a series
 \begin{equation}
\t = x^0 + \int_y A_1^{\mu\nu}(x, y) h_{\mu\nu}(y)+ \frac{1}{2}  \int_{y_1} \int_{y_2} A_2^{\mu\nu\rho\sigma}(x,y_1, y_2) h_{\mu\nu}(y_1) h_{\rho\sigma}(y_2)+ \dots\;,
 \end{equation}
where the coefficients $A_1^{\mu\nu}$, etc., can be determined by plugging the series into \eqref{Curvauture_expansion} and expanding in $h_{\mu\nu}$ \cite{Pons:2009cd,Frob:2017gyj,Frob:2017lnt}. We have fixed the first term to be $x^0$ so that when $h_{\mu\nu} = 0$ we have $\Phi(x) = \phi(x)$.  

Since we know that we can absorb any choice of the measure into a choice of the action, we can set $\Omega_g = \mu$ without loss of generality. This gives a natural choice for $\t$ which amounts to the condition $G^{00} =-1$.

Let us calculate $A_1 = A^{\mu\nu}_1h_{\mu\nu}$ with the higher order coefficients $a_1$, etc. set to zero. One then finds $A_1 = -\frac{1}{2} \int {\rm d}t \, \Theta(x^0-t) h_{00}$, so 
\begin{equation}
\t =  x^0    -\frac{1}{2} \int {\rm d}t \, \Theta(x^0-t) h_{00}(t, \vec{x}) + O(h^2)\;,
\end{equation}
where $\Theta(x^0-t)$ is the Heaviside function.
This transforms as a scalar under linearised diffeomorphisms. This amounts to the transformation
\begin{equation}
    h_{00} \to h_{00} + 2\partial_0 \eps_{0} 
\end{equation}
and
\begin{equation}
    \t \to \t + \eps^{\mu} \partial_\mu\t = \t + \eps^0 + \dots
\end{equation}
as can be checked.

\section{Evaluating the effective action with the minimal measure}
\label{sec:Minimal}
Having discussed the importance of an invariant measure, its connection with the phase space formulation and the minimal generalization to QFT in curved spacetime, we wish to evaluate the one-loop effective action.
We will do it using two different regularisations: the mode $N$-cutoff  \cite{Becker:2020mjl,Becker:2021pwo, Banerjee:2023ztr,Ferrero:2024yvw} and the proper-time (or heat kernel) regularisation \cite{DeWitt:1960fc,Bonanno:2004sy,Glaviano:2024hie,Falls:2024noj}.

In this section we start taking the measure factor to be $\Omega_g = \mu$ leaving the study of the general case to the next section.
 With this choice for the measure factor, we can evaluate the Gaussian path integral \eq{Gamma} according to the definition \eq{MeasureDef}, obtaining
\begin{equation} \label{Gamma_eval}
\Gamma[g] = S[g] + \frac{1}{2}  \Tr \log \left((-\nabla^2 +m^2)/\mu^2\right)\;.
\end{equation}
This expression is divergent and we can be regularised it in different ways \cite{Falls:2024noj}.  Each regularisation introduces a cutoff that must be removed to obtain the physical result while adjusting the bare couplings.

\subsection{$N$-cutoff}
Let us start considering an $N$-cutoff \cite{Becker:2020mjl,Becker:2021pwo,Ferrero:2024yvw}, a UV cutoff set on the number of modes. This amounts to include in the trace only modes with eigenvalues of the Lapacian $\lambda_i \leq \lambda_N$ for some finite $N$. For simplicity, let us specialise to the case where the spectrum of $-\nabla^2$ is discrete. Then we consider a complete set of eigenfunctions $f_{i,\ell}(y)$  of the Laplacian for the metric $g_{\mu\nu}$ such that 
\begin{equation}
\sum_i^\infty \sum_r^{c_i} f_{i,r}(x)  \bar{f}_{i,r}(y) = \delta(x-y)   \end{equation}
and
\begin{equation}
\int_x \bar{f}_{i,r}(x)  f_{j,s}(x) = \delta_{ij}\delta_{r s}\;.  
\end{equation}
Here $\lambda_i$ are the eigenvalues of the $-\nabla^2$ and $c_i$ are the multiplicities.
Then we see that the trace can be written as
\begin{eqnarray}
   \Tr \log \left((-\nabla^2 +m^2)/\mu^2\right) &\equiv& \int_x  \lim_{y\to x} \log \left((-\nabla^2 +m^2)/\mu^2\right) \delta(x-y)\\
   && \sum_i^\infty \sum_r^{c_i} \int_x \bar{f}_{i,r}(x) \log \left((-\nabla^2 +m^2)/\mu^2\right) f_{i,r}(x)  \\
  && =  \sum_{i=0}^{\infty}  c_i \log \left((\lambda_i +m^2)/\mu^2\right) \;.
\end{eqnarray}
Since this is divergent, it is not a convergent series.  However, we can check that our formal demonstration of diffeomorphism invariance in the last section translates to the fact that each approximant \cite{Becker:2020mjl}
\begin{equation}
2 \delta S_N = \Tr_N \log \left((-\nabla^2 +m^2)/\mu^2\right) \equiv \sum_{i=0}^{N}  c_i \log \left((\lambda_i +m^2)/\mu^2\right)
\end{equation}
 is diffeomorphism invariant for $N < \infty$. To see this, we note that it follows from 
\begin{equation}
-\nabla^2_g f_{i,r}(x) = \lambda_i f_{i,r}(x) \,,
\end{equation}
for that Laplacian of a metric $g$, that
\begin{eqnarray}
    -\nabla^2_{g'} f_{i,r}'(x) &=&  \int_{\hat{x}} \delta(\hat{X}(x),\hat{x}) (-\nabla^2_g) f_{i,r}(\hat{x}) \\
    &=&  \int_{\hat{x}} \delta(\hat{X}(x),\hat{x})\lambda_i f_{i,r}(\hat{x}) \\
     &=&  \lambda_i f_{i,r}' \,.
\end{eqnarray}
Hence, for each eigenfunction on spacetime with metric $g_{\mu\nu}$ we have a function with the same eigenvalue on the space with metric $g'_{\mu\nu}$. Furthermore, since diffeomorphisms are invertible, the converse is true also, and hence the metrics have the same spectrum. Therefore we have 
\begin{equation}
    \delta S_N[g'] = \delta S_N[g] 
\end{equation}
Thus, at least when we use the $N$-cutoff we see that the formally diffeomorphism-invariant measure gives rise to diffeomorphism-invariant effective action.

\subsection{Proper-time cutoff}
Alternatively we use a proper-time cutoff. This is based on the  regularisation of the trace  (up to terms independent of the metric) via 
\begin{equation} \label{PTregInv}
\frac{1}{2} \Tr \log\left(( - \nabla^2+m^2)/\mu^2 \right) \to -\frac{1}{2} \int_{1/\Lambda^2}^\infty \frac{ds}{s}  \Tr  e^{-(- \nabla^2+m^2)s }  \;,
\end{equation}
where $s$ is the proper time or heat kernel time.
Requiring  the effective action $\Gamma$ in \eqref{Gamma_eval} to be independent of $\Lambda$, a UV cutoff, we then obtain the proper-time flow \cite{Bonanno:2004sy}:
\begin{equation}
    \Lambda \partial_\Lambda S_\Lambda = \Tr[ e^{-\left(- \nabla^2+m^2\right)/\Lambda^2}]\;.
\end{equation}
For large $\Lambda$ the trace can be evaluated using the early time heat kernel expansion \cite{Benedetti:2010nr,Groh:2011dw}. This results in the following flow equation
\begin{eqnarray} \label{flow_diff_inv}
     \Lambda \partial_\Lambda S_\Lambda &=&  \int \sqrt{g} \Big[ \Lambda^d + \frac{1}{6} R \Lambda^{d-2} - m^2 \Lambda^{d-2} + \frac{\Lambda^{d-4}}{2} m^4 - \frac{\Lambda^{d-4}}{6} R m^2 + \frac{\Lambda^{d-4}}{72} R^2 - \frac{\Lambda^{d-4}}{180} R_{\mu\nu}R^{\mu\nu}
     \nn &&+ \frac{\Lambda^{d-4}}{180} R_{\mu\nu\rho\lambda} R^{\mu\nu\rho\lambda}  + O(\Lambda^{d-6}) \Big]\;,
\end{eqnarray}
 which is manifestly diffeomorphism invariant. 
In conclusion, also the proper-time regularisation with $\Omega_g = \mu$ generates a diffeomorphism-invariant effective action.

\section{General ultra-local measure}
\label{sec:General}
Let us now generalize to a general measure factor $\Omega_g$.
With a general measure factor the one-loop effective action is given by
\begin{equation} \label{GammaOmega}
\Gamma[g] = S[g]  + \frac{1}{2}  \Tr \log \left(-\Omega_g^{-2} \nabla^2  +  \Omega_g^{-2} m^2 \right)\;,
\end{equation}
which is not diffeomorphism invariant unless $\Omega_g(x)$ is a scalar. This should be obvious to the undeluded reader since overwise $-\Omega_g^{-2} \nabla^2  +  \Omega_g^{-2} m^2$ is not a covariant operator. Nonetheless, measures have been proposed which correspond to the measure factors \cite{Fradkin:1973wke}
\begin{equation} \label{M_FV}
\Omega_g = \sqrt{g^{00}}
\end{equation}
and \cite{Donoghue:2020hoh}
\begin{equation} \label{M_JD}
\Omega_g = \sqrt{(g)^{-1/d}}\;, 
\end{equation}
neither of which are scalars.

To make it more clear, we evaluate the proper-time flow for a generic $\Omega_g$.

\subsection{Proper-time flows}
We now calculate the divergent part of effective action for a general measure factor $\Omega_g$. 
To this end we UV regularise  
\begin{equation} \label{PTreg}
\frac{1}{2} \Tr \log( - \Omega_g^{-2} \nabla^2) \to -\frac{1}{2} \int_{1/\n^2}^\infty \frac{ds}{s}  \Tr  e^{\Omega_g^{-2} \nabla^2 s }  
\end{equation}
where $\n$ is a cutoff with mass dimension $[\n]=1- [\Omega_g]$.
In the limit $\n\to \infty$ we obtain the unregulated form\footnote{To obtain the previous proper-time regularisation \eq{PTregInv} we can set either $\Omega_g = \mu$, send $s \to s \mu$ and send $\n  \to \Lambda/\mu$ or set simply set $\Omega_g =1$. }.

From \eq{PTreg} we obtain a flowing action $S_\n$ by requiring $\Gamma$ to be independent of $\n$:
\begin{equation} \label{Dimensionless_Flow}
\n \partial_\n S_\n =  \Tr  \exp\left(\Omega_g^{-2} \nabla^2 \n^{-2}\right) \,.
\end{equation}
The trace can be evaluated again for large $\Lambda$ using heat kernel methods, obtaining 
\begin{eqnarray}
 \n \partial_\n S_\n &=&  \int_x  \frac{\sqrt{g}}{ (4\pi)^{d/2}} \Bigg[  \Omega_g^{d} \n^d +\left( \frac{1}{6}   \Omega ^{d-2} \left(R-6 m^2\right)-\frac{1}{12} (d-4) (d-2)
  \Omega ^{d-4} g^{\mu\nu} \partial_\mu \Omega \partial_\nu \Omega  \right) \n^{d-2}
\end{eqnarray}
 The terms of order $\n^{d-4}$ are given in equation \eq{fullflow} of the appendix, where all the details of the calculation of $\n \partial_\n S_\n$ can be found.  
We see that unless $\Omega_g$ is a scalar, the flow generates terms which are not diffeomorphism invariant. With \cite{Branchina:2024lai,Branchina:2024xzh,Fradkin:1973wke,Unz:1985wq} $\Omega_g = \sqrt{g^{00}}$  we see this explicitly 
\begin{eqnarray} \label{eq:flowwithg00}
\n \partial_\n S_\n &=& \frac{1}{ (4\pi)^{d/2}} \int_x \sqrt{g} \Bigg[  \left(g^{00}\right)^{\frac{d}{2}} \n^d +\left( \frac{1}{6} R - m^2 \right)\left(g^{00}\right)^{\frac{d-2}{2}} \n^{d-2}  \nn
&\,&   -\frac{1}{48} (d-2) (d-4) \left(g^{00}\right)^{-2} g^{\mu\nu} \partial_\mu g^{00} \partial_\nu g^{00} \left(g^{00}\right)^{\frac{d-2}{2}} \n^{d-2}  + O(n^{d-4}) \Bigg] \;,
\end{eqnarray}
clearly showing that the choice of measure is not invariant.
Equally, with \cite{Donoghue:2020hoh} $\Omega_g = g^{-1/(2d)}$ we have
\begin{eqnarray}
\n \partial_\n S_\n &=& \frac{1}{ (4\pi)^{d/2}} \int_x \Bigg[   \n^d +\left(  -\frac{1}{48d^2} (d-2) (d-4) g^{-2} g^{\mu\nu} \partial_\mu g \partial_\nu g  + \frac{1}{6} R - m^2 \right)g^{\frac{1}{d}} \n^{d-2}  \nn
&\,&  + O(\n^{d-4}) \Bigg] \,.
\end{eqnarray}
In this case we note that the term proportional to $\Lambda^d$ is trivially invariant since it does not depend on the metric. This is the property that is the focus of \cite{Donoghue:2020hoh}. However, the other terms break diffeomorphism invariance. In particular, we have a term $\int_x g^{\frac{1}{d}} R$ instead of the invariant term  $\int_x \sqrt{g} R$. 

We note that the logarithmic divergencies in all even dimensions are independent of $\Omega_g$ and diffeomorphism invariant. Thus, the measure factor only affects power law divergencies. For example if we take $d=4$ we have 
\begin{eqnarray} \label{flow_diff_inv}
     \Lambda \partial_\Lambda S_\Lambda &=&  \frac{1}{ (4\pi)^{2}} \int_x \sqrt{g} \Big[ \Omega_g^4 \Lambda^4 + \frac{1}{6} R \Lambda^{2} \Omega_g^2 - m^2 \Lambda^{2} \Omega_g^2 + \frac{1}{2} m^4 - \frac{1}{6} R m^2 + \frac{1}{72} R^2 - \frac{1}{180} R_{\mu\nu}R^{\mu\nu}
     \nn &&+ \frac{1}{180} R_{\mu\nu\rho\lambda} R^{\mu\nu\rho\lambda}  + O(\Lambda^{d-6}) \Big]\;.
\end{eqnarray}

As a further step, one can specify the flow equation on a given class of background, in order to do the matching between the LHS and the RHS of the flow. We will show in the next subsection, how this turns out to be a delicate procedure.

\subsection{Constant curvature background}
In  \cite{Branchina:2024xzh} the measure with $\Omega_g = \sqrt{g^{00}}$ was chosen and the authors used a spherical constant curvature background in $d=4$ dimensions. Then the curvature monomials can be re-expressed as
\begin{eqnarray}
\label{eq:R}
  R &=& \frac{12}{a^2}  \\
  R_{\mu\nu}R^{\mu\nu} &=& \frac{36}{a^4} \\
  R_{\mu\nu\rho\lambda} R^{\mu\nu\rho\lambda} &=& \frac{24}{a^4} \\
  \label{eq:Vol}
 \int_x \sqrt{g} &=&  \frac{8 \pi ^2 a^4}{3}
  \end{eqnarray}
  where $a$ is the radius of the sphere.
To determine $\Omega_g = \sqrt{g^{00}}$, \cite{Branchina:2024xzh}  specified that
\begin{equation}  \label{conformal_a}
g_{\mu\nu} = a^{2} \tilde{g}_{\mu\nu}\;,
\end{equation}
where $\tilde{g}_{\mu\nu}$ a metric which is independent of $a$, the unit-sphere metric. This implies in particular that the time-time component
\begin{equation}
g^{00} = a^{-2} \tilde{g}^{00}\,.
\end{equation}
An example of the metric $\tilde{g}_{\mu\nu}$ is 
\begin{equation} \label{tildemetric}
\tilde{g}_{\mu\nu} dx^\mu dx^{\nu} =d\chi^2 + \sin^2 \chi \left( d\theta^2 + \sin^2\theta \, d\phi^2 + \sin^2\theta \sin^2\phi \, d\psi^2 \right)\;.
\end{equation}

Here, some observations are in order. Firstly, there are coordinate systems where \eq{conformal_a} does not hold and the following results are restricted to those coordinate systems for which this holds.   Secondly, we observe that in \cite{Branchina:2024xzh} a term equal to
\begin{equation}
\mathcal{C} = - \log \left(\prod_x \sqrt{\tilde{g}^{00}(x)}\right)
\end{equation}
was not included in the regulated trace and so appears there unregulated. If we use e.g. \eq{tildemetric} (with $x^0 = \chi$) we have that 
\begin{equation}
\tilde{g}^{00}(x)=1 \implies g^{00}(x) = a^{-2}
\end{equation}
and then $\mathcal{C} = 0$.  Then by evaluating \eq{flow_diff_inv} with $\Omega_g = a^{-1}$ and  using \eq{eq:R}-\eq{eq:Vol}, we recover the result of \cite{Branchina:2024xzh}: 
\begin{equation} \label{Branchina}
    \n \partial_\n S_\n = \frac{1}{6} \Lambda^4 +  \left(\frac{1}{3} - \frac{a^2 m^2}{6}\right)\Lambda^2 + \frac{29}{90} - \frac{a^2 m^2}{3} + \frac{a^4 m^4}{12}
\end{equation}
Integrating the flow \eq{Branchina} wrt. the cutoff $\Lambda$, we obtain the regulated effective action
 \begin{equation} \label{Branchina1}
  S_{\Lambda} = \frac{1}{24} \Lambda^4 +  \left(\frac{1}{6} - \frac{a^2 m^2}{12}\right)\Lambda^2 + \frac{29}{180} \log(\Lambda^2) - \frac{a^2 m^2}{6} \log(\Lambda^2)  + \frac{a^4 m^4}{24}  \log(\Lambda^2) +  \text{ constant}\,.
 \end{equation}
With $S^{\rm eff}_{\rm grav} = -S_{\Lambda} +  \text{ constant}$ and $\Lambda = N$  this agrees with \cite{Branchina:2024xzh} (c.f. equation (51) of that paper).

However, we can use another coordinate system for a constant curvature metric and, since the measure has broken diffeomorphism invariance, we will find a different result. For example, we can take
\begin{equation} \label{DeSitter}
g_{\mu\nu} dx^{\mu} dx^{\nu}  = dt^2 + a^2 \cos^2\left(\frac{t}{a}\right) \left[ d\theta^2 + \sin^2\theta \left( d\phi^2 + \sin^2\phi \, d\psi^2 \right) \right]\,,
\end{equation}
(from which we obtain de Sitter space by sending $t \to {\rm i} t$). 
Then from \eq{DeSitter} we have that 
\begin{equation}
g^{00} =1
\end{equation}
and we obtain 
\begin{equation}\label{eq:good}
    \n \partial_\n S_\n = \frac{a^4}{6} \Lambda^4 + \left(\frac{a^2}{3} - \frac{a^4 m^2}{6}\right) \Lambda^2 + \frac{29}{90} - \frac{a^2 m^2}{3} + \frac{a^4 m^4}{12}\;,
\end{equation}
which is equal to the result computed with an invariant measure, i.e. when we evaluate  \eq{flow_diff_inv} on a four sphere for either  \eq{conformal_a},  \eq{DeSitter} or any metric for the four sphere. We stress $a$ is the radius of the sphere both in \eq{conformal_a} (with  \eq{tildemetric})  and in \eq{DeSitter}. However, we can get different results depending on the coordinate system if we choose a measure that breaks diffeomorphism invariance. This demonstrates the inherent  inconsistency of \cite{Branchina:2024lai,Branchina:2024xzh}.

\section{The measure in quantum gravity}\label{sec:brst}
In order to arrive at the BRST invariant path integral measure for quantum gravity \cite{Fujikawa:1983im}, we can utilize a DeWitt metric for the metric itself $ \gamma^{\mu\nu, \rho\sigma }(x,y)$ and a metric on the space of diffeomorphisms $\eta_{\mu, \nu}(x,y)$. For Einstein gravity, we can take these to be given by 
\begin{equation}
 \gamma^{\mu\nu, \rho\sigma }(x,y) =  \frac{\mu^{2}}{64 \pi G_N}  \sqrt{g} \left(g^{\mu\rho}g^{\nu \sigma} + g^{\mu\sigma}g^{\nu\rho} - \alpha  g^{\mu\nu} g^{\rho \sigma} \right)\delta(x-y)\,,
\end{equation}
where $\alpha$ is the DeWitt parameter\footnote{The most natural choice for this parameter turns out to be $\alpha =1$ since then, at least at one-loop, some factors cancel. Other values will lead to a constant factor.},
and
\begin{equation}\label{eq:eta}
\eta_{\mu, \nu}(x,y) = \frac{\mu^4}{16 \pi G} \sqrt{g} g_{\mu\nu} \delta(x-y)\,,
\end{equation}
respectively. 
The metric transforms under a BRST transformation as
\begin{equation}
    \delta_\theta g_{\mu\nu} = (\nabla_\mu c_{\nu}+ \nabla_\nu c_{\mu}) \theta\;,
\end{equation}
where $\theta$ is a Grassmannian parameter.
Then, consequently, the ghosts $c^{\mu}$, anti-ghosts $\bar{c}_\mu$ and Nakanishi-Lautrup field $b_\mu$  transform as 
\begin{equation}
    \delta_\theta c^{\mu} = - c^{\nu} \nabla_\nu c^{\mu} \theta =- c^{\nu} \partial_\nu c^{\mu} \theta \, \,\,\,\,\,\,\,\,\,\delta_\theta \bar{c}_{\mu} = b_\mu \theta\,,\,\,\, \,\,\,\, \,\delta_\theta b_{\mu} = 0\,.
\end{equation}
Hence, one can check that the measure \cite{Falls:2020tmj}
\begin{equation}
    D( g) D( b) D(c) D (\bar{c}) \frac{\sqrt{\det \gamma[g]}}{\sqrt{\det \eta[g]} }
\end{equation}
is BRST invariant. It takes some work to appreciate this.
First, we note that the $(D\bar{c})$ and $(D b)$ are trivially invariant, so that what has to happen is that transformation of factors depending on the two DeWitt metrics must cancel the nonzero Jacobians we pick up from transforming $(D g) $ and $(D c)$. This conspires because the metrics are BRST symmetric in the sense of possessing ``Killing vectors''.
Indeed the metric  obeys the ``Killing equation''
\begin{equation}
\int_{x,y,z} v_{\mu\nu}(x)  \left( \delta_\theta g_{\a\b}(z)   \frac{\delta \gamma^{\mu\nu, \rho\sigma }(x,y)}{\delta g_{\a \b}(z) }  +  \gamma^{\mu\nu, \alpha \beta }(x,z) \frac{\delta\, \delta_\theta g_{\a\b}(z) }{\delta g_{\rho  \lambda}(y) }     +  \gamma^{\a\b, \rho \lambda  }(x,z) \frac{\delta \, \delta_\theta g_{\a\b}(z)}{\delta g_{\mu\nu}(y) } \right) v_{\rho \lambda}(y) = 0   
\end{equation}
where $v^{\mu\nu}(x)$ is a test field.
By taking the trace of this equation, i.e. contracting indices of the coefficients of the test fields with the inverse DeWitt metric and also integrating,  we have that
\begin{equation}
\delta_\theta \sqrt{\det \gamma[g]} = - \sqrt{\det \gamma[g]} \int_z  \frac{\delta \, \delta_\theta g_{\a\b}(z)}{\delta g_{\a\b}(z) }\;,
\end{equation}
which cancels the Jacobian we get from transforming $(D g)$.  Additionally, we note that 
\begin{equation}
\int_y \frac{\delta \, \delta_\theta c^\mu(x)}{\delta c^\nu(y)} v^\nu(y)  = v^\nu(x) \nabla_\nu c^\mu(x) -  c^\nu(x) \nabla_\nu v^\mu(x)
\end{equation}
from which it follows that $\eta$ $\gamma$ in \eqref{eq:eta} too obeys the ``Killing equation''
\begin{equation}
\int_{x,y,z} v^\mu (x)\left(\delta_\theta g_{\a\b}(z) \frac{\delta \eta_{\mu,\nu}(x,y)}{\delta g_{\a \b}(z)}  +  \eta_{\alpha,\nu}(z,y)   \frac{\delta \, \delta_\theta c^\alpha(z)}{\delta c^\mu(x)}+\eta_{\mu,\a}(x,z)  \frac{\delta \, \delta_\theta c^\a(z)}{\delta c^\nu(y)}\right) v^\nu(y)\;.
\end{equation}
Taking the trace of this equation then leads to
\begin{equation}
    \delta_\theta \frac{1}{\sqrt{\det \eta[g]} } =  \frac{1}{\sqrt{\det \eta[g]} } \int_z  \frac{\delta \, \delta_\theta c^\a(z)}{\delta c^\nu(z)}\;,
\end{equation}
which cancels the Jacobian coming from the transformation of $D( c)$ (remembering that $c$ is Grassmannian). Hence, the measure is BRST invariant.

As with the measure and action of the scalar, we can always define what we mean by the action, by writing
\begin{equation}
\mathcal{Z} = \int  D (g) D (b) D(c) D (\bar{c})  \frac{\sqrt{\det \gamma[g]}}{\sqrt{\det \eta[g]} } {\rm e}^{-\S[g,c,\bar{c}, b]}\;,
\end{equation}
where $\S[g,c,\bar{c}, b]$ should be BRST invariant. 
Then, by convention all the freedom to select a theory is determined by $\S[g,c,\bar{c}, b]$. This form agrees with the measure of \cite{Fujikawa:1983im} up to constants and is the minimal form.

Let us now confront the controversy of factors of $g^{00}$. To do so we can perform a relational quantisation of the theory as we did for a scalar field. One way to do this is to fix that gauge such that \cite{Papertoappear} $\Sigma^{I}(x) \!!! x^{I} $ so that in particular 
\begin{equation} \label{RelationalGauge}
\tau(x) \!!! x^0\;.
\end{equation}
Thus the dependence of the action on $b$ should linear so that it implements a delta function which imposes these conditions. In fact for any gauge condition one can in principle find the corresponding scalars.  With this gauge choice we can proceed which the canonical analysis carried out in \cite{Unz:1985wq} which differs from Fujikawa's by the replacement
\begin{equation}
\mu^2 \to \mu^2 g^{00}\,.
\end{equation}
This leads to \cite{Unz:1985wq}
\begin{equation}
\mathcal{Z} = \int  D(g) D(b) D (c) D( \bar{c})\left( \prod_x \left(\sqrt{g^{00}(x)}\right)^{\frac{d(d-3)}{2}} \right) \frac{\sqrt{\det \gamma[g]}}{\sqrt{\det \eta[g]} } {\rm e}^{-\S[g,c,\bar{c}, b]}\;,
\end{equation}
which is {\it not} BRST invariant. 
However, because of the gauge condition \eq{RelationalGauge}, one has
\begin{equation}
g^{00} \!!! g^{\mu\nu} \partial_\mu\t \partial_\nu \t\;.
\end{equation}
Thus, the situation unfolds as follows.  On the one hand if we use $g^{00}$ in the measure it breaks BRST invariance.  On the other hand it is consistent to replace $g^{00}$ with the the Fradkin-Vilkovisky scalar $g^{\mu\nu} \partial_\mu \t \partial_\nu \t$ under the gauge condition \eq{RelationalGauge}.  Thus we arrive at  
\begin{equation}
\mathcal{Z} = \int  D (g) D( b) D (c) D(\bar{c})\left( \prod_x \left(\sqrt{g^{\mu\nu} \partial_\mu\t \partial_\nu \t}\right)^{\frac{d(d-3)}{2}} \right) \frac{\sqrt{\det \gamma[g]}}{\sqrt{\det \eta[g]} } {\rm e}^{-\S[g,c,\bar{c}, b]}\;,
\end{equation}
which is manifestly BRST invariant. 
At the same time, the factor
\begin{equation}
    \left( \prod_x \left(\sqrt{g^{\mu\nu} \partial_\mu\t \partial_\nu \t }\right)^{\frac{d(d-3)}{2}} \right) = {\rm e}^{\frac{d-3}{4} \Tr \log o}  \,, 
\end{equation}
with
\begin{equation}
 o(x,y) \equiv  g^{\mu\nu}(x) \partial_\mu\t(x) \partial_\nu \t(x) \delta(x-y) \,,
\end{equation}
is by itself diffeomorphism invariant and hence BRST invariant. As such it can be absorbed into the action by sending 
\begin{equation}
\S[g,c,\bar{c}, b] \to \frac{d-3}{4} \Tr \log o + \S[g,c,\bar{c}, b] \;,
\end{equation}
bringing the measure back to Fujikawa's minimal form.

\section{Consequences of a ``bad'' measure for the RG flow of quantum gravity}\label{sec:flow}

As we have seen in detail for the case of a scalar field, a measure that breaks diffeomorphism invariance can have dramatic effects on the RG flow. Indeed, the measure studied in \cite{Branchina:2024lai} breaks diffeomorphism invariance and also leads the authors to the conclusion that there is no asymptotically safe fixed point in quantum gravity. 

The situation in quantum gravity mirrors what we saw for the scalar field and can be summarised as follows. If the invariant minimal measure is used, factors $\mu^2$ appearing in the measure are, through the use of a cutoff, identified with the cutoff scale $\Lambda^2$.
This generically leads to RG flows that feature an RG UV fixed point. In particular one gets terms in the RG flow in four dimensions of the form 
\begin{equation}
   b_0 \Lambda^4 \int_x \sqrt{g}
\end{equation}
and
\begin{equation}
   b_1 \Lambda^2 \int_x \sqrt{g} R\;,
\end{equation}
which are essentially the terms needed for a fixed point for the cosmological constant and Newton's constant. Without such terms the mechanism that usually generates the fixed point is lost.

On the other hand the measure used in  \cite{Branchina:2024lai} instead, exchanges $\mu^2$ with $g^{00}$. If $g^{00}$ is assumed to have dimension of a mass squared, this has the effect of replacing 
\begin{equation}
\Lambda^2  \to g^{00} N^2\;,
\end{equation}
where $N$ is a dimensionless cutoff. This replaces 
\begin{equation}
   b_0 \Lambda^4 \int_x \sqrt{g} \to   b_0 N^4 \int_x \left(g^{00}\right)^2  \sqrt{g}\,,
\end{equation}
\begin{equation}
   b_1 \Lambda^2 \int_x \sqrt{g} R \to  b_1  N^2 \int_x  g^{00}\sqrt{g} R \,.
\end{equation}
For an arbitrary metric this would lead to the rather obvious conclusion that one has broken diffeomorphism invariance. 
However, if, as was assumed in \cite{Branchina:2024lai}, $g^{00} \propto R $ for a particular metric we have (up to a constant)
\begin{equation}
    \Lambda^2 \to N^2 R\,,
\end{equation}
and thus
\begin{equation}
   b_0 \Lambda^4 \int_x \sqrt{g} \to   b_0 N^4 \int_x   \sqrt{g} R^2\,,
\end{equation}
\begin{equation}
   b_1 \Lambda^2 \int_x \sqrt{g} R \to  b_1  N^2 \int_x \sqrt{g} R^2 \,.
\end{equation}
Therefore, apparently the terms normally providing a fixed point for the Newton's constant and the cosmological constant instead appear to renormalise the coupling to $\sqrt{g} R^2$. However, the truth is that they are non-invariant terms reflecting the choice of measure.

Before closing this section, let us consider a BRST invariant choice of measure that could be viewed as the ``covariant version'' of the measure used in \cite{Branchina:2024lai,Branchina:2024xzh}, in the sense that it has the same effect. This is achieved by choosing the Fradkin-Vilkovisky scalar to be
\begin{equation}
\label{tina}
g^{\mu\nu} \partial_\mu\t(x) \partial_\nu \t(x)  = R(x)/\mu^2\,,
\end{equation}
instead of the minimal choice $g^{\mu\nu} \partial_\mu\t(x) \partial_\nu \t(x)  =1$.
This has the same effect on the RG flow as the one observed in \cite{Branchina:2024lai,Branchina:2024xzh} since it leads also to $\Lambda^2 \to N^2 R$.
From the point of view of symmetry, this choice is acceptable. However, the measure now contains a factor that vanishes in flat spacetime.  On the other hand (\ref{tina}) is certainly pathological, as it would fail to hold or become ill-defined for hyperbolic manifolds, or more generally, for Ricci-flat spaces.

\section{Conclusions}\label{sec:conclusion}
In this paper, we aim to clarify both the choice of the appropriate measure and its contested role in determining the renormalisability of the theory.

With respect to the choice of the ``correct measure'', we would like to clarify the following: while there is indeed some freedom in defining the measure from the canonical phase space path integral, whichever consistent approach one takes the measure is unique up to terms that can be absorbed into a diffeomorphism-invariant action. 
Nonetheless, this apparent ambiguity sparked the debate between the approaches of Fradkin and Vilkovisky \cite{Fradkin:1973wke} versus Fujikawa \cite{Fujikawa:1983im} and Toms \cite{Toms:1986sh}. The debate was primarily centered on the diffeomorphism invariance (or BRST invariance) of the theory and the manner in which the  time-time component of the inverse metric $g^{00}$  is incorporated into the phase space variables.
Therefore, the debate centers around the question of  whether the factor \eq{eq:factor} breaks diffeomorphism invariance or not.

In this paper, we address this question directly.
In particular, we have seen explicitly that, when the factors of $g^{00}$ are included in the measure, the resulting effective action is not  invariant  under diffeomorphism. The flow of the Wilsonian effective action \eq{eq:flowwithg00} makes this explicit.
Moreover, we illustrate that, despite the inherent freedom in choosing the measure, a minimal prescription can be established, resulting in Fujikawa's measure. We detailed how to arrive at this scheme beginning with the examples quantum mechanics, QFT in flat Minkowski case, and  QFT in curved spacetime. A final step to quantum gravity follows the same spirit with BRST symmetry as the guide.
While there is the freedom to go beyond the minimal form, there is no room for non-covariant factors of $g^{00}$ in a BRST symmetric construction. Instead, covariant factors $g^{\mu\nu} \partial_\mu \t \partial_\nu \t$ are permissible, but can be absorbed into the action.

Concerning implications for the non-perturbative renormalisation of gravity along the lines of asymptotic safety, we should emphasise that one searches for the correct theory and does not propose measures which are supposed to be ``correct''. On the other hand proposals such as \cite{Branchina:2024lai,Donoghue:2020hoh} that break diffeomorphism invariance and seem to disfavor a fixed point seem doubly ill motivated.   

\bigskip

We anticipate that our analysis will become increasingly valuable in ``relational'' approaches quantum gravity that seek a gauge invariant description and, more broadly, in the study of non-perturbative quantum gravity. Indeed, apart from the approach of asymptotic safety, one could attempt a first-principles approach that specifies a form of the path integral starting from the canonical theory. Our message for such approaches is that the choice of measure can be parameterized by the Fradkin-Vilkovisky scalar. While one can specify a preferred choice of gauge invariant variables,  one could also specify the scalar and reverse engineer the gauge variables.
One can even envisage an important role that the measure may play in connecting different approaches to quantum gravity \cite{Thiemann:2024vjx,Ferrero:2024rvi} and its effects concerning open issues such as unitarity, locality and renormalisability.

\section*{Acknowledgments}
The authors are grateful to Damiano Anselmi, Roberto Percacci, Martin Reuter and Frank Saueressig  for fruitful  discussions and Frank Saueressig and Roberto Percacci for comments on the manuscript. A.B. is grateful to Giuseppe Di Fazio for fruitful discussions on proper-time representation of elliptic operators.\\
R.F.  is supported by the  Friedrich-Alexander University (FAU) Emerging Talents Initiative (ETI).  R.F. is grateful for the hospitality of Perimeter Institute
where part of this work was carried out.
Research at Perimeter Institute is supported in part by the Government of Canada
through the Department of Innovation, Science and Economic Development and by the
Province of Ontario through the Ministry of Colleges and Universities. This work was
supported by a grant from the Simons Foundation (grant no. 1034867, Dittrich).\\
\\

\begin{appendix}
\section{Heat kernel coefficients under a conformal transformation}\label{app:details}
In this appendix we will report the  computation of the heat kernel coefficients resulting from a conformal transformation. These coefficients will particularly useful for the results in section \ref{sec:General}.
For calculation purposes, it is useful to define a new metric 
\begin{equation}
\tilde{g}_{\mu\nu} = \Omega_g^2(x) g_{\mu\nu}(x)\;,
\end{equation}
which is a tensor if $\Omega_g$ is a scalar.
Therefore to have a consistent notation we treat $\Omega_g(x)$ as a scalar. Importantly
{\it the meaning of the covariant derivative on $\Omega_g(x)$ therefore assumes it is a scalar.}

When $\nabla^2$ acts on a scalar we have that 
\begin{equation}
\nabla^2 = \frac{1}{\sqrt{g}} \partial_\mu ( \sqrt{g} g^{\mu\nu} \partial_\nu) =   \frac{1}{\sqrt{g}} \partial_\mu ( \sqrt{g} g^{\mu\nu} \partial_\nu)= \frac{\Omega_g^d}{\sqrt{\tilde{g}}} \partial_\mu ( \Omega_g^{2-d} \sqrt{\tilde{g}} \tilde{g}^{\mu\nu} \partial_\nu) =  \Omega_g^2 \tilde{\nabla}^2 -(d-2) \tilde{g}^{\mu\nu} \Omega_g \partial_\mu \Omega_g \partial_\nu
\end{equation}
so (in expressions with tildes we use the tilde metric to raise and lower indices) 
\begin{equation}
-\Omega_g^{-2} \nabla^2 = - \tilde{\nabla}^2 + (d-2) \tilde{g}^{\mu\nu} \Omega_g^{-1} \partial_\mu \Omega_g \partial_\nu = -\tilde{\nabla}^2 - 2\tilde{A}^\mu\tilde{\nabla}_\mu\;,
\end{equation}
where we defined the vector $\tilde A^\mu$ consistently as
\begin{equation}
\tilde{A}^{\mu} = - \frac{1}{2}  (d-2)  \tilde{g}^{\mu\nu} \Omega_g^{-1} \partial_\nu \Omega_g = - (d-2)  \tilde{\nabla}^\mu \log( \Omega_g)\;. 
\end{equation}
Next, we define a new covariant derivative \cite{Benedetti:2010nr} as $\tilde{D}_\mu = \tilde{\nabla}_\mu + \tilde{A}_\mu$
such that
\begin{equation}
\tilde{\Delta} =  -\tilde{D}^2 + \tilde{E} =-\Omega_g^{-2}\nabla^2 + \Omega_g^{-2} m^2 
\end{equation}
is a minimal Laplacian-type operator,
where the endomorphism can be identified with
\begin{equation}
\tilde{E} = \tilde{A}_\mu \tilde{A}^{\mu} +  \tilde{\nabla}^\mu \tilde{A}_{\mu} +  \Omega_g^{-2} m^2 
\end{equation}
The ``curvature" associated to $D_\mu$ is zero
\begin{equation}
\tilde{F}_{\mu\nu} = [\tilde{D}_\mu,\tilde{D}_\nu]=   \tilde{\nabla}_\mu \tilde{A}_{\nu} - \tilde{\nabla}_\nu \tilde{A}_{\mu} = 0
\end{equation}
Hence, we have that the oneloop effective action \eq{GammaOmega} is given by
\begin{equation}
\Gamma[g]= \frac{1}{2} \Tr \log ( - \tilde{D}^2 + \tilde{E})
\end{equation}
or equivalently
\begin{equation}
\Gamma[g]= \frac{1}{2} \Tr \log ( \tilde{\Delta} )   \,,\,\,\,\,\,\,\,\,    \tilde{\Delta} =  - \tilde{D}^2 + \tilde{E}\;.
\end{equation}

In order to evaluate the flow equation in the presence of a non-trivial measure, we can compute the heat kernel coefficients associated to $\tilde \Delta$.
The proper-time flow equation with a dimensionless cutoff is given by 
\begin{equation} \label{Dimensionless_Flow}
\n \partial_\n S_\n =  \Tr  \exp\left(\tilde{\Delta} \n^{-2}\right) \;.
\end{equation}
We can then  evaluate the LHS via the heat kernel coefficients up to order $\tilde B_2$
\begin{equation} \label{tildetrace}
\Tr\exp\left(\tilde{\Delta} \n^{-2}\right) =  \frac{1}{ (4\pi)^{d/2}} \int_x \sqrt{\tilde{g}} \left[ \tilde{B}_0 \n^d +\tilde{B}_1 \n^{d-2}  + \tilde{B}_2 \n^{d-4} \dots \right]
\end{equation}
with
\begin{equation}
 \tilde{B}_0  =1\;,
\end{equation}
\begin{equation}\label{eq:B1}
\tilde{B}_1 = - \tilde{E}  + \frac{1}{6} \tilde{R} \;,
\end{equation}
\begin{equation}\label{eq:B2}
\tilde{B}_2 = \frac{1}{2} \tilde{E}^2 - \frac{1}{6} \tilde{R} \tilde{E} + \frac{1}{72} \tilde{R}^2 - \frac{1}{180} \tilde{R}_{\mu\nu} \tilde{R}^{\mu\nu} + \frac{1}{180} \tilde{R}_{\mu\nu\alpha\beta} \tilde{R}^{\mu\nu \alpha\beta}\;.
\end{equation}
Expressing the trace in \eq{tildetrace} in terms of the original metric one can explicitely appreciate which differences there are and which new terms get generated.
Using that the transformed curvature is \cite{Wald:1984rg}
\bea
  \tilde{R}
 &=&   \Omega_g^{-2} (R - 2 (d-1)  \Omega_g^{-1} \nabla^2 \Omega_g  -(d-4) (d-1) \Omega_g^{-2} \nabla^\mu \Omega_g \nabla_\mu \Omega_g)\;,
\eea
the first coefficient \eqref{eq:B1} becomes
\begin{equation}
\int_x \sqrt{\tilde{g}} \tilde{B}_1 = \int_x \sqrt{g}\left( \frac{1}{6}   \Omega ^{d-2} \left(R-6 m^2\right)-\frac{1}{12} (d-4) (d-2)
  \Omega ^{d-4} \nabla_\mu \Omega \nabla^\mu \Omega  \right)
   \end{equation}
The computation of the transformed \eqref{eq:B2} is a bit more involved and gives
\bea
     \int_x \sqrt{\tilde{g}} \tilde{B}_2   &=& \int_x \sqrt{g} \Bigg[ \frac{1}{360} (d-6) (5 d-16) (\nabla^2 \Omega)^2 \Omega ^{d-6}-\frac{1}{180} (d-6)
   (d-2) \nabla_\mu \nabla_\nu \Omega \nabla^\mu  \nabla^\nu \Omega \Omega ^{d-6}\nn
   && +\nabla_\mu \Omega \nabla^\mu \Omega \left(\frac{1}{360} (d
   (5 d-36)+60) R \Omega ^{d-6}-\frac{1}{12} (d-4)^2 m^2 \Omega
   ^d\right)-\frac{1}{45} (d-6) R^{\mu\nu} \nabla_\mu  \Omega  \nabla_\nu \Omega \Omega ^{d-6} \nn
   &&+\frac{\frac{1}{360} (d-6) (d (5 d-38)+68) \nabla^2 \Omega \nabla_\mu \Omega \nabla^\mu \Omega
   \Omega ^d+\frac{1}{45} (d-6) (d-2) \nabla^\mu  \Omega \nabla_\mu \nabla_\nu \Omega \nabla^\nu \Omega \Omega ^d}{\Omega
   ^7} \nn && +\frac{\nabla^2 \Omega \left(\frac{1}{180} (5 d-18) R \Omega ^d-\frac{1}{6}
   (d-4) m^2 \Omega ^d\right)+\frac{1}{90} (d-6) R^{\mu\nu} \nabla_\mu \nabla_\nu
   \Omega ^d}{\Omega ^5} \nn && +\frac{(d-6) (d (d (5 d-58)+196)-224) (\nabla_\mu \Omega \nabla^\mu \Omega)^2
   \Omega ^{d-8}}{1440} \nn &&+\frac{\frac{m^4 \Omega ^d}{2}-\frac{1}{6}
   m^2 R \Omega ^d+\frac{R^2 \Omega ^d}{72}-\frac{R_{\mu\nu} R^{\mu\nu} \Omega
   ^d}{180}+\frac{R_{\mu\nu\rho\lambda} R^{\mu\nu\rho\lambda} \Omega ^d}{180}}{\Omega ^4}   \Bigg]
\eea
Summarizing, plugging all these coefficients into the flow equation in \eqref{Dimensionless_Flow} one gets
\bea
\label{fullflow}
\n \partial_\n S_\n &=& \frac{1}{ (4\pi)^{d/2}} \int_x \sqrt{g} \Bigg[  \Omega_g^{d} \n^d +\left( \frac{1}{6}   \Omega ^{d-2} \left(R-6 m^2\right)-\frac{1}{12} (d-4) (d-2)
  \Omega ^{d-4} \nabla_\mu \Omega \nabla^\mu \Omega  \right) \n^{d-2} \nn
&&+  \Bigg( \frac{1}{360} (d-6) (5 d-16) (\nabla^2 \Omega)^2 \Omega ^{d-6}-\frac{1}{180} (d-6)
   (d-2) \nabla_\mu \nabla_\nu \Omega \nabla^\mu  \nabla^\nu \Omega \Omega ^{d-6}\nn
   && +\nabla_\mu \Omega \nabla^\mu \Omega \left(\frac{1}{360} (d
   (5 d-36)+60) R \Omega ^{d-6}-\frac{1}{12} (d-4)^2 m^2 \Omega
   ^d\right)-\frac{1}{45} (d-6) R^{\mu\nu} \nabla_\mu  \Omega  \nabla_\nu \Omega \Omega ^{d-6} \nn
   &&+\frac{\frac{1}{360} (d-6) (d (5 d-38)+68) \nabla^2 \Omega \nabla_\mu \Omega \nabla^\mu \Omega
   \Omega ^d+\frac{1}{45} (d-6) (d-2) \nabla^\mu  \Omega \nabla_\mu \nabla_\nu \Omega \nabla^\nu \Omega \Omega ^d}{\Omega
   ^7} \nn && +\frac{\nabla^2 \Omega \left(\frac{1}{180} (5 d-18) R \Omega ^d-\frac{1}{6}
   (d-4) m^2 \Omega ^d\right)+\frac{1}{90} (d-6) R^{\mu\nu} \nabla_\mu \nabla_\nu
   \Omega ^d}{\Omega ^5} \nn && +\frac{(d-6) (d (d (5 d-58)+196)-224) (\nabla_\mu \Omega \nabla^\mu \Omega)^2
   \Omega ^{d-8}}{1440} \nn &&+\frac{\frac{m^4 \Omega ^d}{2}-\frac{1}{6}
   m^2 R \Omega ^d+\frac{R^2 \Omega ^d}{72}-\frac{R_{\mu\nu} R^{\mu\nu} \Omega
   ^d}{180}+\frac{R_{\mu\nu\rho\lambda} R^{\mu\nu\rho\lambda} \Omega ^d}{180}}{\Omega ^4} \Bigg) \n^{d-4}  \Bigg]\,.
\eea

\end{appendix}

\end{document}